\documentclass{caosp306}

\usepackage{graphicx}

\usepackage{natbib}
\articleNo{123}
\pubyear{2005}
\volume{35}
\volnumber{3}
\firstpage{1}
\received{May 1, 2005}
\accepted{August 28, 2005}

\def\BibTeX{{\rm B\kern-.05em{\sc i\kern-.025em b}\kern-.08em
             T\kern-.1667em\lower.7ex\hbox{E}\kern-.125emX}}

\begin{document}

%
\hauthor{M. Berton et al.}

\title{Line shapes in narrow-line Seyfert 1 galaxies: a tracer of physical properties?}


%
\author{
       M. Berton \inst{1,2}
      \and 
        I. Bj\"orklund \inst{3}   
	\and
       A. L\"ahteenm\"aki \inst{2,3}
	\and
	E. Congiu \inst{4}
	\and
	E. J\"arvel\"a \inst{5,2}
	\and
	G. Terreran\inst{6}
	\and
	G. La Mura \inst{7}
       }

\institute{
           Finnish Centre for Astronomy with ESO (FINCA), University of Turku, Quantum, Vesilinnantie 5, FI-20014, University of Turku, Finland, \email{marco.berton@utu.fi}
         \and 
          Aalto University Mets{\"a}hovi Radio Observatory, Mets{\"a}hovintie 114, FI-02540 Kylm{\"a}l{\"a}, Finland 
         \and 
           Aalto University Department of Electronics and Nanoengineering, P.O. Box 15500, FI-00076, Aalto, Finland 
	 \and
		Las Campanas Observatory - Carnegie Institution of Washington, Colina El Pino Casilla 601, La Serena, Chile
	\and
		Department of Physics, University of California, Santa Barbara, CA 93106-9530, USA
	\and
	Center for Interdisciplinary Exploration and Research in Astrophysics CIERA, Department of Physics and Astronomy, Northwestern University, Evanston, IL 60208, USA
	\and
	Laboratory of Instrumentation and Experimental Particle Physics, Av. Prof. Gama Pinto, 2 - 1649-003 Lisboa, Portugal
          }

\date{March 8, 2003}

\maketitle

\begin{abstract}
Line profiles can provide fundamental information on the physics of active galactic nuclei (AGN). In the case of narrow-line Seyfert 1 galaxies (NLS1s) this is of particular importance since past studies revealed how their permitted line profiles are well reproduced by a Lorentzian function instead of a Gaussian. This has been explained with different properties of the broad-line region (BLR), which may present a more pronounced turbulent motions in NLS1s with respect to other AGN. We investigated the line profiles in a recent large NLS1 sample classified using SDSS, and we divided the sources into two subsamples according to their line shapes, Gaussian or Lorentzian. The line profiles clearly separate all the properties of NLS1s. Black hole mass, Eddington ratio, [O III], and Fe II strength are all very different in the Lorentzian and Gaussian samples. We interpret this in terms of evolution within the class of NLS1s. The Lorentzian sources may be the youngest objects, while Gaussian profiles may be typically associated to more evolved objects. Further detailed spectroscopic studies are needed to fully confirm our hypothesis.
\end{abstract}

\newcommand{\kms}{km s$^{-1}$}
\newcommand{\ergs}{erg s$^{-1}$}
\newcommand{\Hb}{H$\beta$}
\newcommand{\chired}{$\chi^2_\nu$}
\section{Introduction}

Among active galactic nuclei (AGN), narrow-line Seyfert 1 galaxies (NLS1s) are often considered as ``rebels", which hardly fit into the unified models \citep{Antonucci93, Urry95}. Recognized as a separate class more than 30 years ago \citep{Osterbrock85}, by definition they are those AGN in which the full width at half maximum (FWHM) of H$\beta$ is lower than 2000 \kms\ and the flux ratio [O III]$\lambda$5007/H$\beta < 3$ \citep{Osterbrock85}. An additional criterion for NLS1s is the presence of strong Fe II multiplets \citep{Goodrich89}, although their presence is not always required, and the origin of the iron abundance and its physics are still not clear (\citealp{Marziani18b}, Bon et al., 2019, submitted). Their steep X-ray spectral indexes \citep{Boller92} and the presence of relativistic jets in these AGN \citep{Abdo09a, Abdo09c, Berton18c} further complicate the NLS1 picture. \par
Some of these characteristics can be explained, from a physical point of view, by the presence of a relatively low-mass black hole (10$^6$-10$^8$ M$_\odot$, see \citealp{Peterson11}) which is accreting close to or above the Eddington limit \citep{Boroson92, Boller96, Sulentic00, Marziani01, Zhou06, Cracco16, Rakshit17a, Chen18}. Due to this, NLS1s are located at the extreme end of the so-called eigenvector 1 (EV1), and constitute a class of extreme accretors \citep{Sulentic15, Sniegowska18, Marziani18b}, a property which would allow a more physically accurate definition than the classic FWMH-based criterion \citep{Marziani18a}. \par
An explanation which accounts for the low black hole mass, the high accretion rate, and several other NLS1 properties, is that these AGN are younger with respect to classical broad-line Seyfert 1 (BLS1s, e.g., see \citealp{Mathur00, Berton16c, Berton17, Jarvela17})\footnote{\footnotesize Unless otherwise specified, in the following we will use the term BLS1s referring to both non-jetted sources, like classical Seyfert 1, and jetted sources with broad permitted lines such as high-excitation radio galaxies and flat-spectrum radio quasars.}. In this phase, these AGN are growing fast to eventually turn into fully-developed BLS1s. Although this point is still debated \citep[e.g., see][]{Dammando18, Sbarrato18}, much evidence has been accumulated in support of the evolutionary hypothesis (for full reviews on their multiwavelength properties, see \citealp{Komossa08a, Gallo18, Komossa18, Lister18}). \par
Interestingly enough, a property which seems to differ in NLS1s and BLS1s is the profile of their permitted lines, particularly that of H$\beta$. While most type 1 AGN typically have a Gaussian line profile, in NLS1s the profile is instead well reproduced by a Lorentzian function \citep{Moran96, VeronCetty01, Sulentic02, Cracco16}. This is particularly true when only the broad component of \Hb\ is considered. Most broad-line profiles indeed can be well fitted by a broad component, either Gaussian or Lorentzian, and a weaker narrow Gaussian component. \par
However, it is known that a small fraction of NLS1s have Gaussian line profiles. Whether the different line profile is connected to different physical properties is still unclear. To answer this question, we decided to study the new large sample of NLS1s selected from the Sloan Digital Sky Survey (SDSS) by \citet[][hereafter R17]{Rakshit17a}. In this proceeding we will carry out a simple statistical analysis of the line profiles, connect them with the other observed properties of NLS1s, and provide a physical interpretation for these results based on the evolutionary hypothesis. In Sect.~2 we present the sample selection, in Sect.~3 we discuss the impact of redshift on our sample, in Sect.~4 we present our results, in Sect.~5 we discuss them, and in Sect.~6 we provide a brief summary of this work. In the following, we adopt the standard $\Lambda$CDM cosmology, with H$ = 70$ km s$^{-1}$ Mpc$^{-1}$, $\Omega_m = 0.3$, $\Omega_\Lambda = 0.7$ \citep{Komatsu11}. 

\section{Sample selection}

\begin{figure}[!t]
\begin{minipage}[t]{\textwidth}
\includegraphics[width=0.5\hsize,clip=]{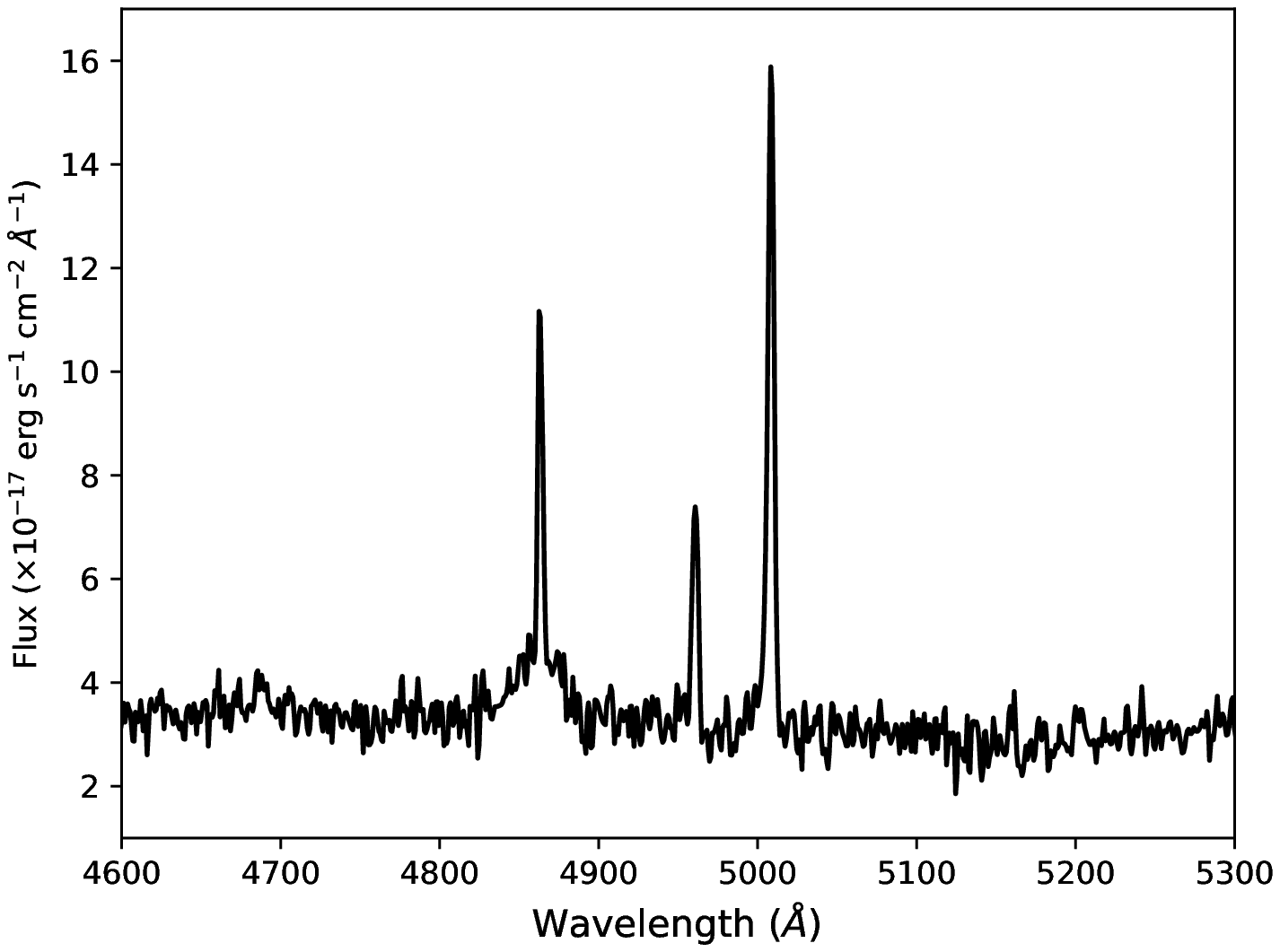}
\includegraphics[width=0.5\hsize,clip=]{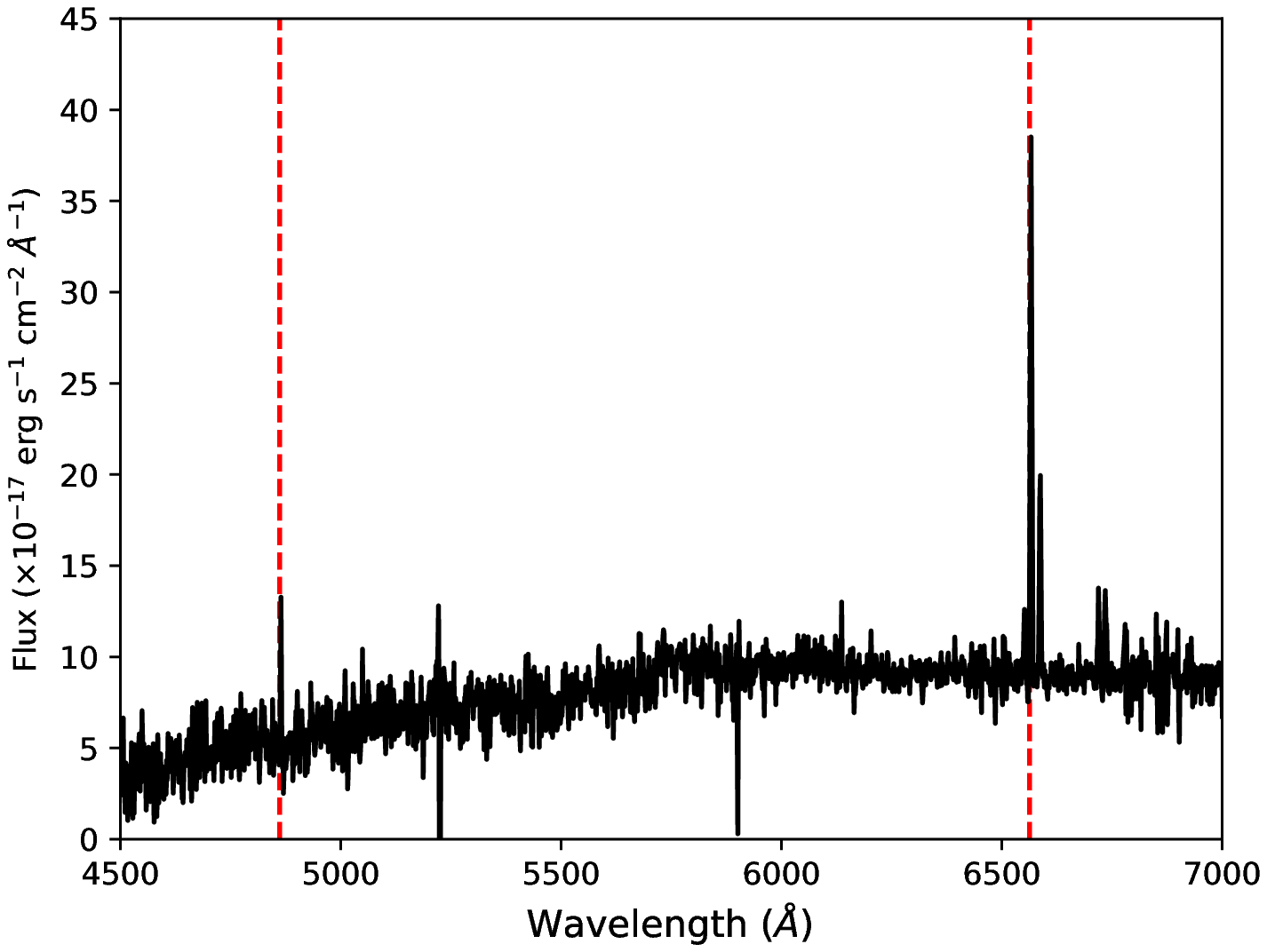}
\end{minipage}
\caption{Two examples of spurious sources present in R17. \textbf{Left panel:} the spectrum of SDSS J143900.17+431440.9 (spec-6061-56076-0704). The three lines are, from left to right, H$\beta$, [O III] $\lambda$4959, [O III] $\lambda$5007. Although classified as an NLS1, this object is an intermediate type AGN, as shown by the presence of two well-separated kinematic components in the H$\beta$ profile. \textbf{Right panel:} the spectrum of SDSS J155806.47+523748.2 (spec-0618-52049-0008), which can be classified as a LINER. The NLS1 classification was likely due to a wrong redshift reported in SDSS (z = 0.401 in SDSS, while the correct value is 0.069). The lines marked with the vertical red dashed lines are H$\beta$ and H$\alpha$, which were used to measure the correct redshift value. }
\label{fig:bad}
\end{figure}

As mentioned above, we used the sample of 11101 NLS1s selected by R17. They selected the sample starting from the SDSS DR12, using the standard classification criteria of FWHM(\Hb) and [O III]/\Hb\ flux ratio. The redshift limit of R17 is z$<$0.8. Among the sources they selected, we considered only those whose signal-to-noise ratio (SNR) in the $\lambda$5100\AA\ continuum is higher than 5, finding 7070 sources. The SNR criterion decreases the chances of finding sources where an appropriate spectral analysis cannot be performed. As shown by \citet{Sulentic15}, spectral measurement can often be unreliable in noisy spectra. Indeed, in R17 some spurious non-NLS1 sources are definitely present, in particular intermediate type AGN, which R17 did not consider but meet both the FWHM and [O III]/H$\beta$ criteria of NLS1s. Examples of an intermediate Seyfert and of a wrong identification are shown in Fig.~\ref{fig:bad}. \par
On the remaining 7070 sources, we performed an extremely simple test. The profile of permitted lines in type 1 AGN typically shows two kinematic components, physically corresponding to the broad-line region (BLR) and the narrow-line region (NLR). However, in the NLS1 population the \Hb\ line usually has an extremely weak narrow component \citep[see Fig. 3 by][]{Sulentic02}. The line profile, after redshift correction and continuum subtraction, can indeed be fitted with a single function to reproduce the BLR emission. Therefore, we fitted each line profile with a Lorentzian and Gaussian profiles and used the reduced chi-squared \chired\ test to select the best fit. We included only sources which are well-reproduced with either of these simple models to our final sample. \par
By means of visual inspection of the spectra and of all the results of the fitting procedure, we decided to apply the following selection criteria also on the $\chi^2_\nu$. We excluded all those sources for which the $\chi^2_\nu \leq 0.5$ and $\chi^2_\nu \geq 5$, because in most cases their spectra were either noisy or misclassified as shown in Fig.~\ref{fig:bad}. The lower limit in $\chi^2_\nu$ was introduced to avoid overfitting. When 3.5$<$\chired$<$5, we accepted only sources with SNR$>$20. Only in high SNR objects, in fact, such high \chired\ still indicates a good representation of the H$\beta$ profile. Between 2.5$<$\chired$<$3.5, we also imposed the criterion SNR$>$15, while if 2$<$\chired$<$2.5 we imposed SNR$>$10. All remaining sources with 0.5$<$\chired$<$2 were included in the samples. To decide whether a source is to be included in the Lorentzian (L) or the Gaussian (G) subsample, we considered the minimum distance from a \chired=1. These selection criteria inevitably miss some NLS1s. However, our aim is not to obtain complete subsamples, but to include a statistically significant number of sources both in G and L sample, and selecting at the same time only genuine NLS1s. Using this technique, we found 3933 NLS1s whose line profile can be reproduced with a single component, either Gaussian or Lorentzian. In particular, 2894 of them exhibit a Lorentzian profile (L sources), while the remaining 1039 sources are better reproduced with a single Gaussian profile (G sources). \par
We recognize the simplicity of this model. Firstly, we did not correct for the Galactic absorption, because on the very short wavelength interval we are considering (4000-5500\AA) it is basically a constant and it can be neglected. We also did not subtract the host galaxy contribution and the Fe II multiplets. However, our procedure selects those sources where a single component is a good representation of the H$\beta$ line. This means that if the host galaxy contribution and the Fe II pseudo-continuum cause a significant deviation from an extremely simple line profile, the source was rejected and not included in the sample because of our selection criteria. In conclusion, this technique automatically selects sources in which Fe II and host galaxy do not significantly contribute to the H$\beta$ profile. A more detailed analysis of the other line profiles, including an accurate subtraction of host galaxy and Fe II components, will be published in an upcoming paper (Berton et al., in prep.). \par
These factors not accounted for the line fitting would have a much stronger influence in measurements of the physical quantities. However, all the properties we are using are extracted from the R17 catalog, in which the Galactic absorption, the host galaxy, and the Fe II contribution have been taken into account. For each source, we used the measurements of redshift, R4570\footnote{R4570 is defined as the flux ratio between the Fe II multiplets and the H$\beta$ line.}, black hole mass, Eddington ratio, and [O III]$\lambda$5007 luminosity. For each property we carried out a Kolmogorov-Smirnov (K-S) test to compare the distributions of Lorentzian and Gaussian sources. In this case, the null hypothesis is that the two distributions of sources originate from the same population of objects. The p-values of each test are shown in the last column of Table~\ref{tab:stats}. When the distributions are not normal but log-normal like, we also tested the differences between their tails by applying the Anderson-Darling (A-D) test \citep{Hou09, Berton16b}. The null hypothesis, again, is that the two distributions originate from the same population. In both cases we rejected the null hypothesis only at a 99\% confidence level, thus when the p-value is lower than 0.01. \par

\begin{figure}[!t]
\begin{minipage}[t]{\textwidth}
\includegraphics[width=0.5\hsize,clip=]{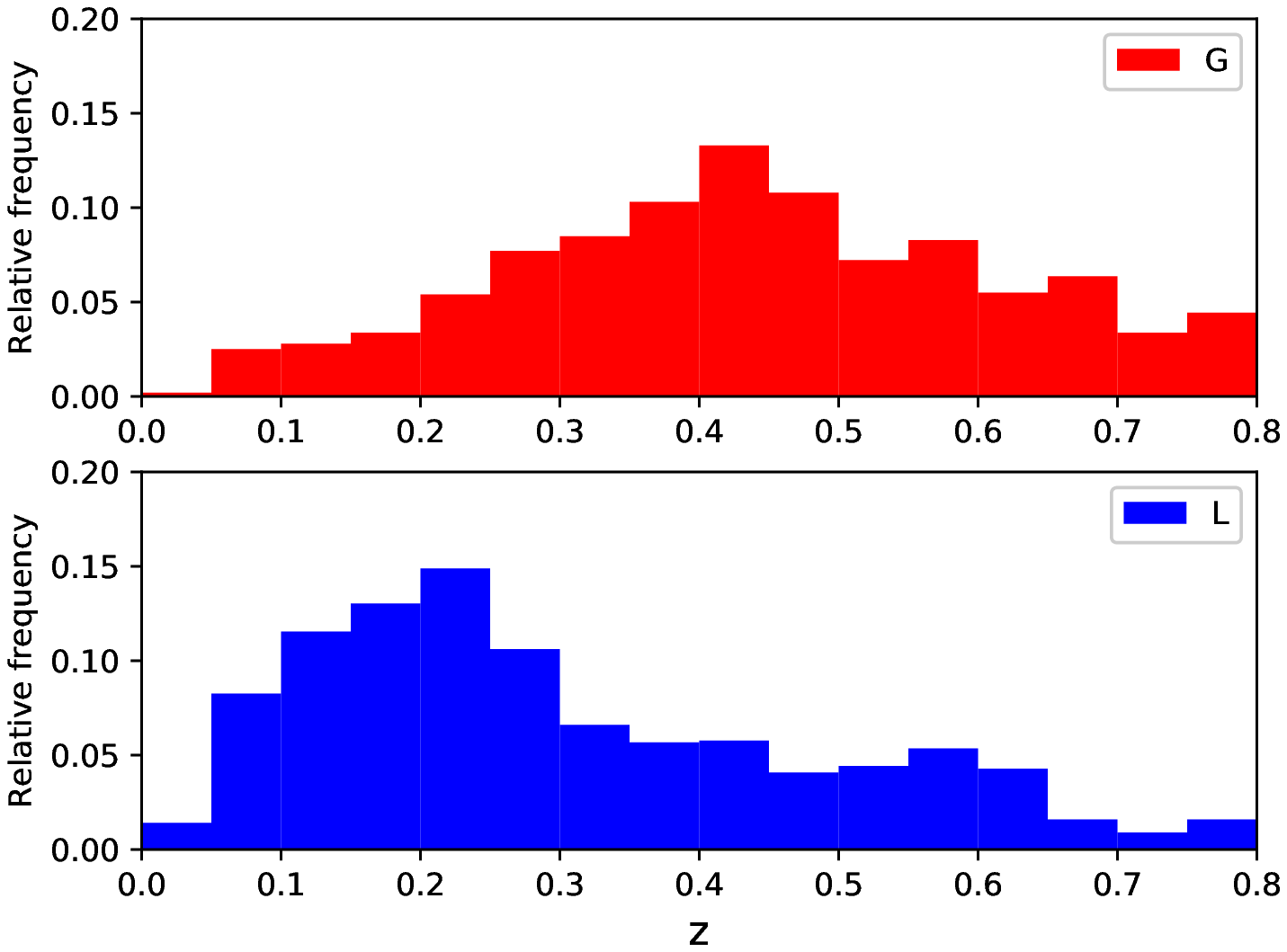}
\includegraphics[width=0.5\hsize,clip=]{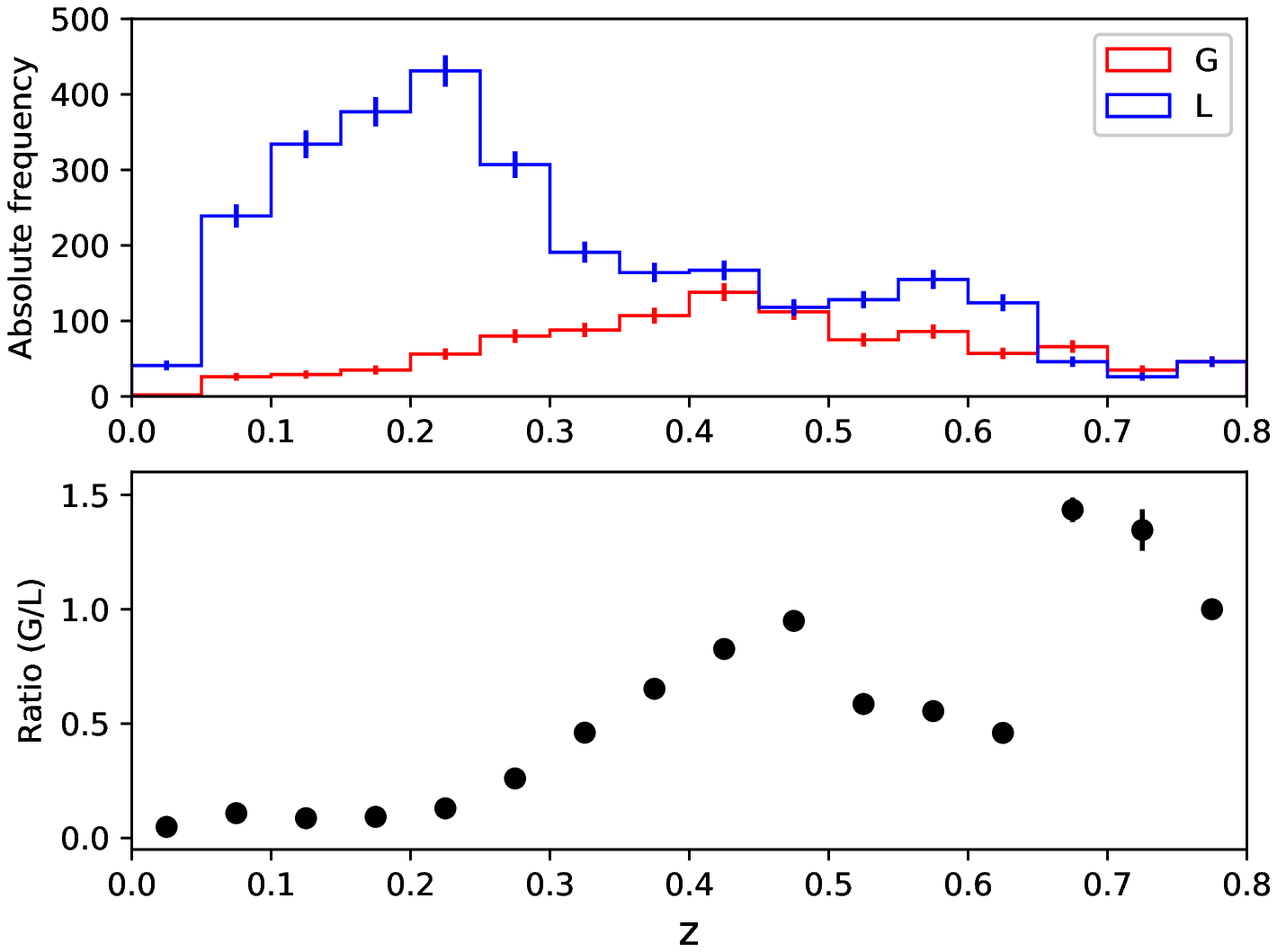}
\end{minipage}
\caption{Redshift distributions of the two samples. The red color represents sources with a Gaussian profile, while the blue color indicates Lorentzian profiles. In the left figure we show the normalized distribution of sources per redshift bin. Hereafter each redshift bin has a width of 0.05. In the right figure, the top panel shows the total number of sources per redshift bin, while the bottom panel shows the ratio between the number of G and L sources as a function of redshift. }
\label{fig:histo_z}
\end{figure}

\section{Redshift distributions}
\label{sec:redshift}
The first result is that the redshift distribution of G and L sources are very different (p-value 1$\times10^{-97}$). The normalized frequencies of the distributions are shown in the left panel of Fig.~\ref{fig:histo_z}, while in the right panel we show the total number of L and G sources per redshift bin, and the ratio between G and L sources. The error associated to each bin assumes a Poisson distribution, so it is the square root of the number of sources in each bin. The ratio remains roughly constant up to $z$ = 0.25, with approximately $\sim$10\% of G sources and 90\% of L sources. This result is comparable to the results by \citet{Cracco16}, who found that up to $z$ = 0.35 the fraction of L sources is 96\%. The small difference between these result can be accounted to different selection criteria. \citet{Cracco16}, in fact, selected only sources with very high SNR, and this could introduce a small bias toward luminous objects. \par
Above $z$ = 0.25, the number of Gaussian sources tends to increase rapidly, while the number of L objects decreases. Above $z$ = 0.5 there are more sources in the G sample than in the L sample. This result could be attributed to a selection effect. When the redshift increases, the SNR tends to decrease, so the wings typical of Lorentzian profiles tend to disappear in the continuum noise. Therefore, without prominent wings a Lorentzian profile can be easily modeled with a Gaussian. We tried to test the effect of redshift on line profiles by comparing it with the SNR ratio. In fact, a selection effect should be present if SNR systematically decreases with redshift. \par
Indeed, we found that SNR has a non-negligible dependence on the redshift, with a Pearson correlation coefficient of $-$0.54 (p-value $\sim$0), and a Spearman rank of $-$0.58 (p-value $\sim$0). Therefore, redshift may play a role, even if it may not be the only factor causing a the difference between Lorentzian and Gaussian sources. In fact, we note that Gaussian sources are found also when the SNR is high. Therefore, the difference between these two samples does not seem to be solely due to a selection effect. However, to make our results more robust we decided to test the distributions of physical properties of G and L sources not only on the entire sample, but also at z$<$0.25, where the ratio between L and G sources is roughly constant and redshift effects are very likely negligible. 

\begin{figure}[!t]
\begin{minipage}[t]{\textwidth}
\includegraphics[width=0.5\hsize,clip=]{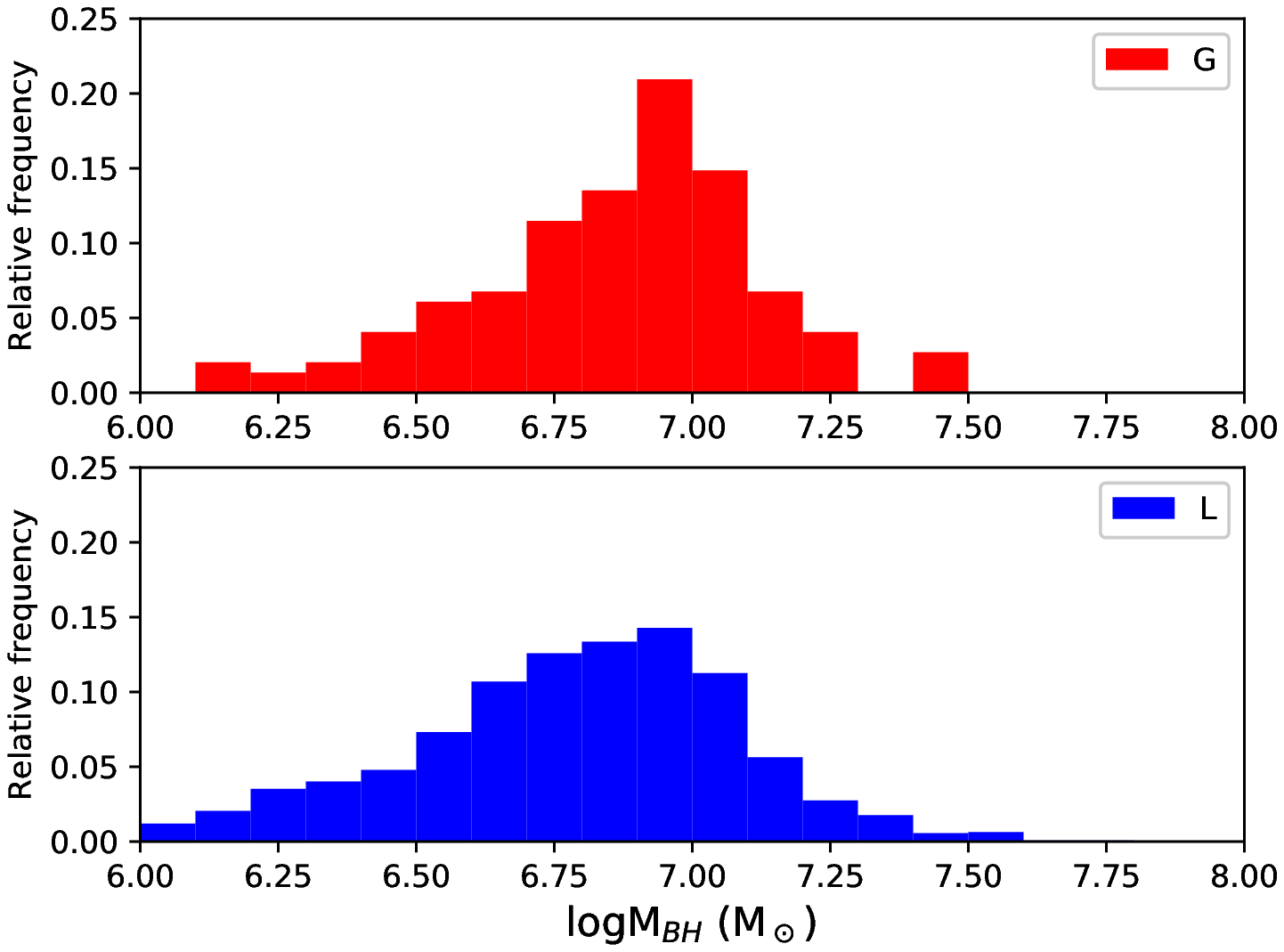}
\includegraphics[width=0.5\hsize,clip=]{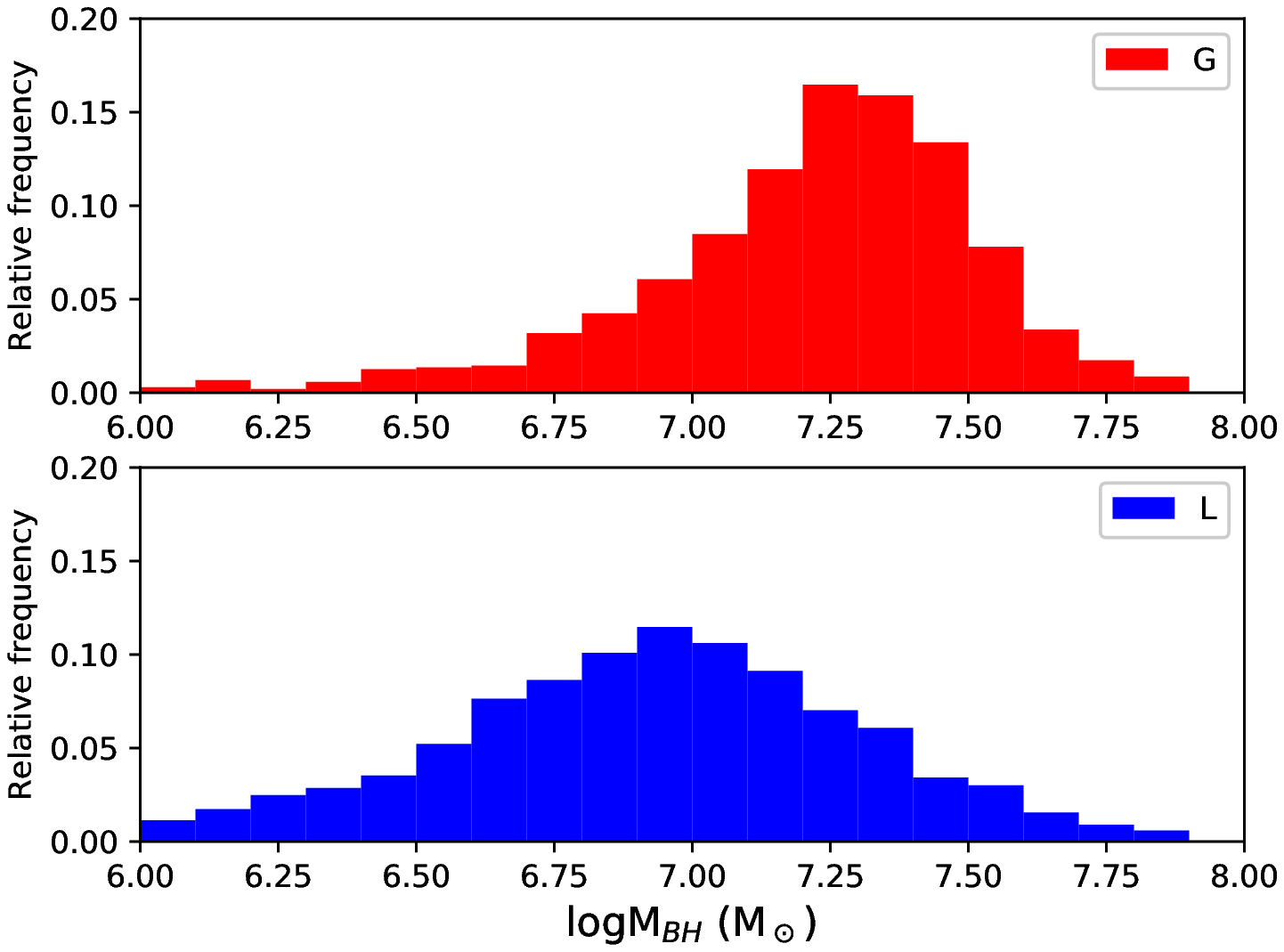}
\end{minipage}
\caption{Distribution of black hole masses of the two samples. \textbf{Left panel:} distribution of sources up to the redshift limit z$=$0.25; \textbf{right panel:} distributions of the whole samples, with redshift limit z$=$0.8. Colors as in Fig.~\ref{fig:histo_z}.}
\label{fig:histo_bh}
\end{figure}

\begin{figure}[!t]
\begin{minipage}[t]{\textwidth}
\includegraphics[width=0.5\hsize,clip=]{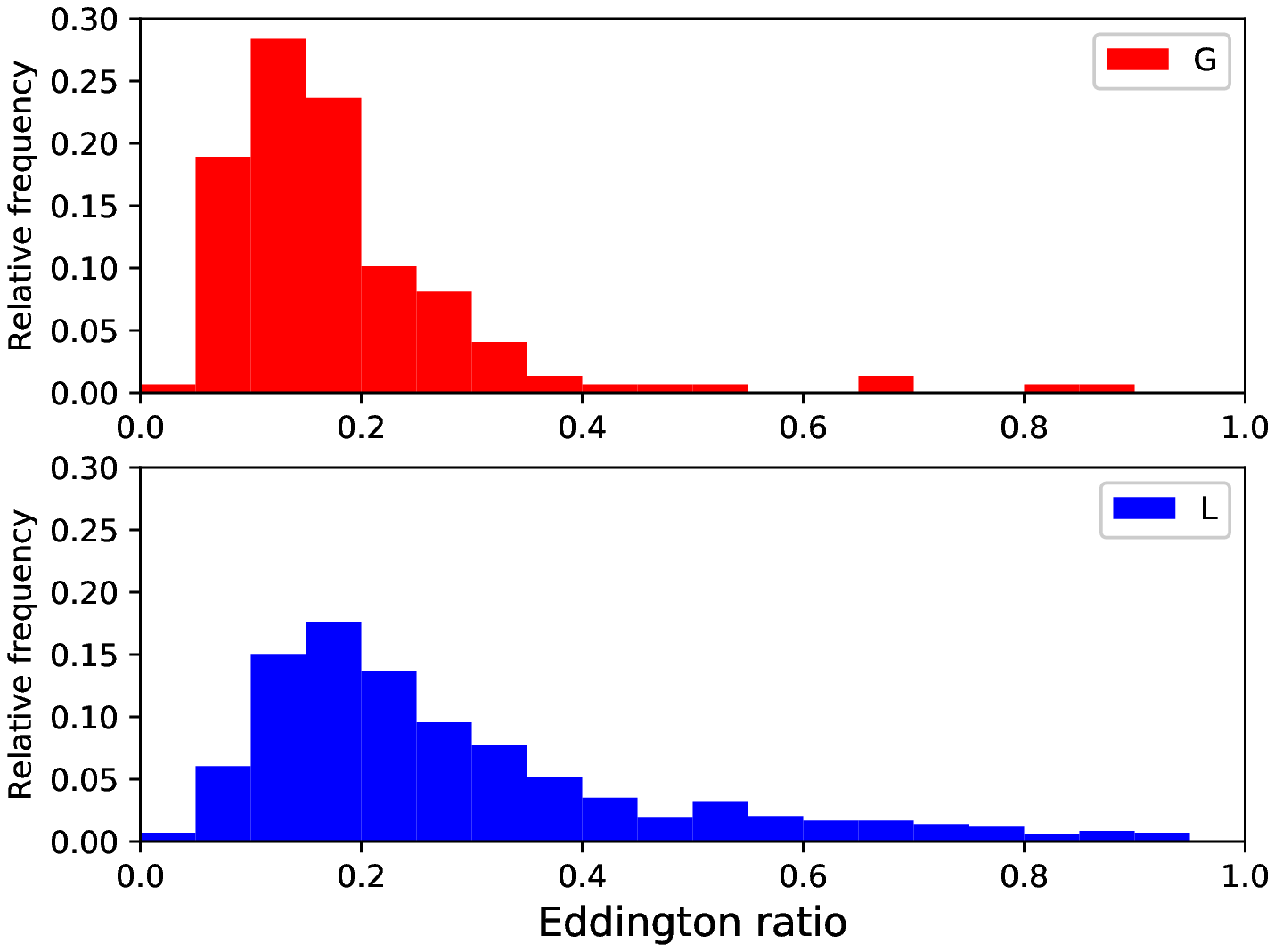}
\includegraphics[width=0.5\hsize,clip=]{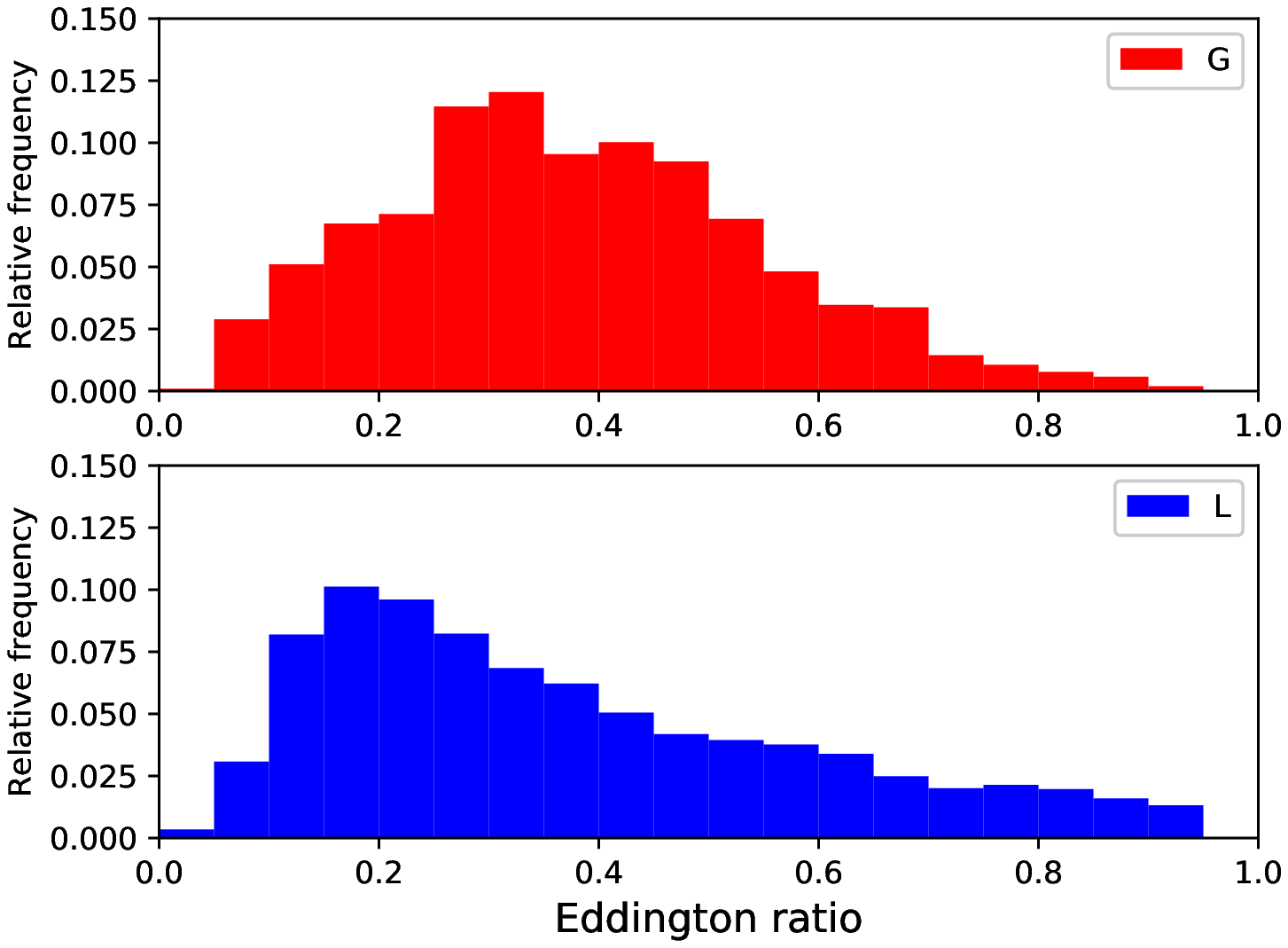}
\end{minipage}
\caption{Distribution of Eddington ratio of the two samples. Colors as in Fig.~\ref{fig:histo_z} and panels as in Fig.~\ref{fig:histo_bh}.}
\label{fig:histo_edd}
\end{figure}

\begin{table}
\caption{Statistical properties of the samples. Columns: (1) measured property (black hole mass in $M_\odot$, Eddington ratio and R4570 adimensional, logarithm of the monochromatic continuum luminosity ($\lambda$L$_\lambda$) at 5100 \AA\ in \ergs, and logarithm of the [O III] luminosity in erg s$^{-1}$,); (2) mean in the Gaussian sample; (3) median in the Gaussian sample; (4) standard deviation in the Gaussian sample; (5) mean in the Lorentzian sample; (6) median in the Lorentzian sample; (7) standard deviation in the Lorentzian sample; (8) K-S test p-value.}
\label{tab:stats}
\centering
\begin{tabular}{l | c c c | c c c | c}
\hline\hline
Property & Mean & Median & $\sigma$ & Mean & Median & $\sigma$ & p-value\\
\hline\hline
z $<$ 0.8 & & G & & & L &  \\
\hline
$\log$M$_{BH}$ & 7.21 & 7.26 & 0.32 & 6.90 & 6.93 & 0.41 & 1$\times$10$^{-111}$ \\
Edd ratio & 0.43 & 0.38 & 0.41 & 0.60 & 0.38 & 0.75 & 4$\times$10$^{-21}$ \\
R4570 & 0.62 & 0.59 & 0.37 & 0.67 & 0.62 & 0.40 & 2$\times$10$^{-3}$ \\
$\log \lambda$L$_{\rm{cont}}$ & 43.92 & 44.0 & 0.48 & 43.66 & 43.66 & 0.53 & 1$\times$10$^{-49}$ \\
$\log$L$_{\rm{[O III]}}$ & 41.54 & 41.55 & 0.47 & 41.35 & 41.34 & 0.52 & 5$\times$10$^{-22}$ \\
\hline\hline
z $<$ 0.25 & & G & & & L & \\
\hline
$\log$M$_{BH}$ & 6.82 & 6.90 & 0.32 & 6.75 & 6.80 & 0.35 & 5$\times$10$^{-3}$ \\
Edd ratio & 0.19 & 0.16 & 0.13 & 0.35 & 0.24 & 0.36 & 3$\times$10$^{-14}$ \\
R4570 & 0.60 & 0.60 & 0.42 & 0.63 & 0.56 & 0.40 & 2$\times$10$^{-1}$ \\
$\log \lambda$L$_{\rm{cont}}$ & 43.18 & 43.20 & 0.43 & 43.33 & 43.34 & 0.42 & 5$\times$10$^{-4}$ \\
$\log$L$_{\rm{[O III]}}$ & 42.11 & 42.11 & 0.42 & 41.84 & 41.88 & 0.70 & 2$\times$10$^{-11}$ \\
\hline
\end{tabular}
\end{table}

\begin{figure}[!h]
\begin{minipage}[t]{\textwidth}
\includegraphics[width=0.5\hsize,clip=]{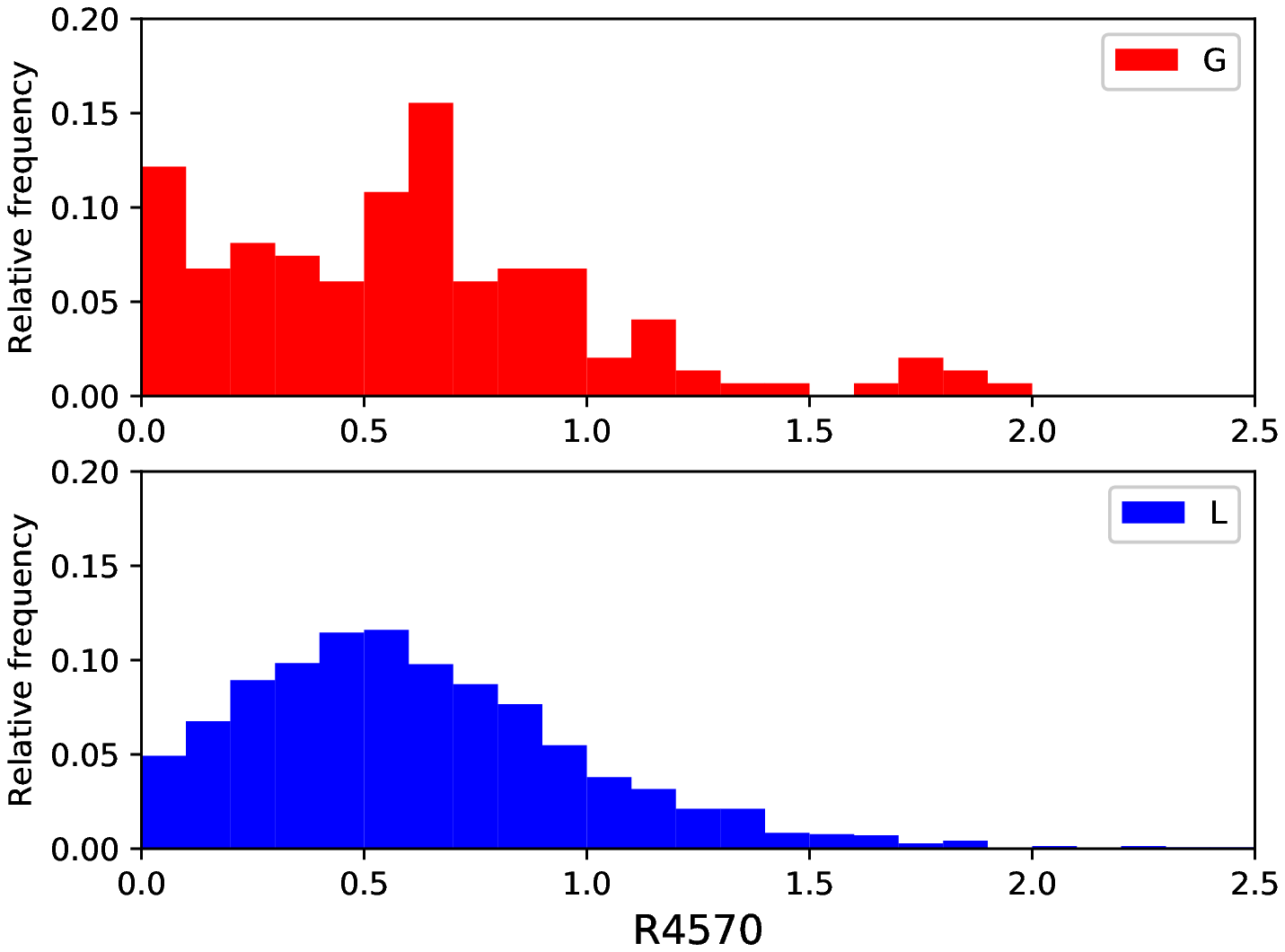}
\includegraphics[width=0.5\hsize,clip=]{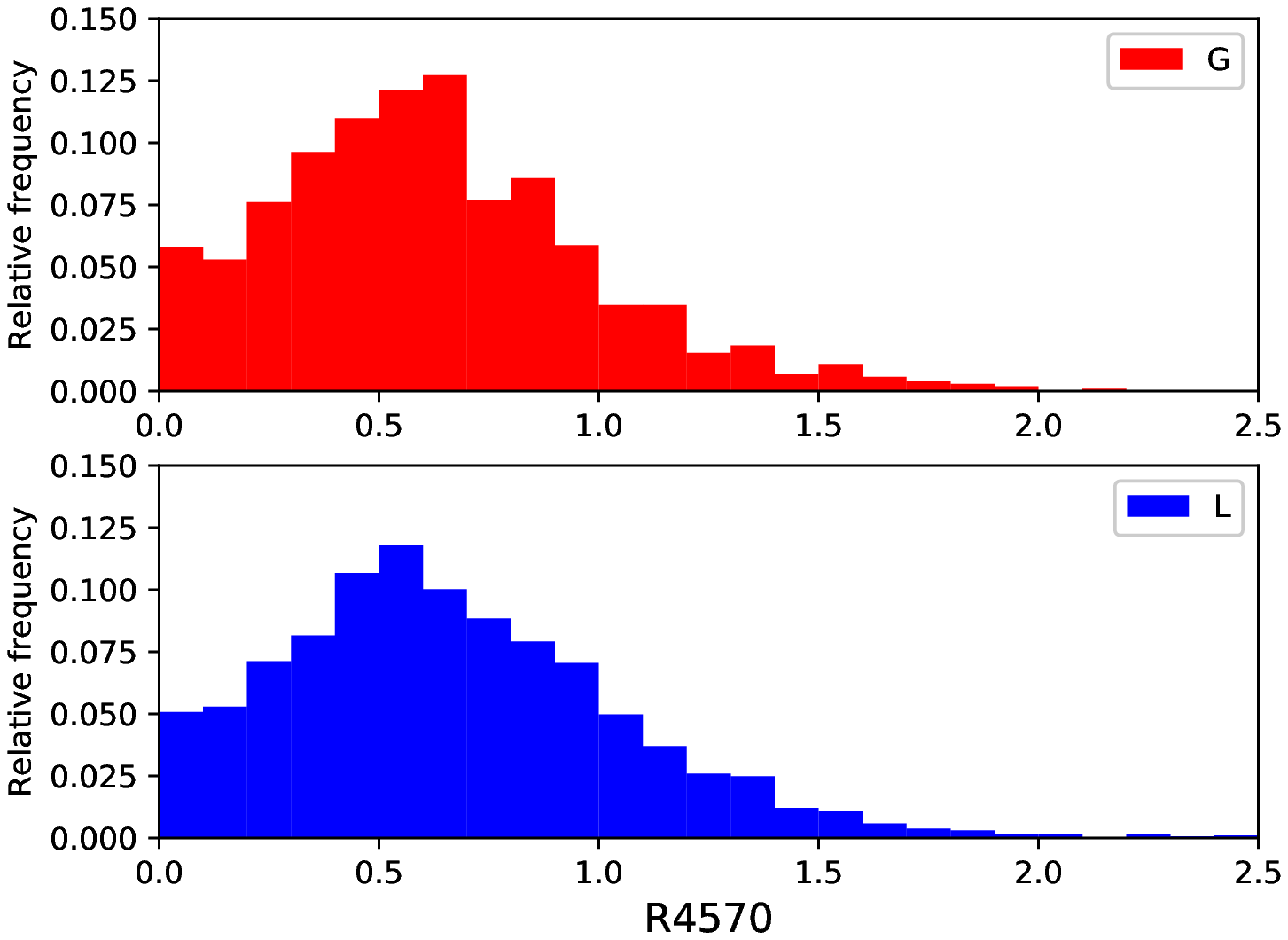}
\end{minipage}
\caption{Distribution of R4570 of the two samples. Colors as in Fig.~\ref{fig:histo_z} and panels as in Fig.~\ref{fig:histo_bh}.}
\label{fig:histo_fe}
\end{figure}

\begin{figure}[!h]
\begin{minipage}[t]{\textwidth}
\includegraphics[width=0.5\hsize,clip=]{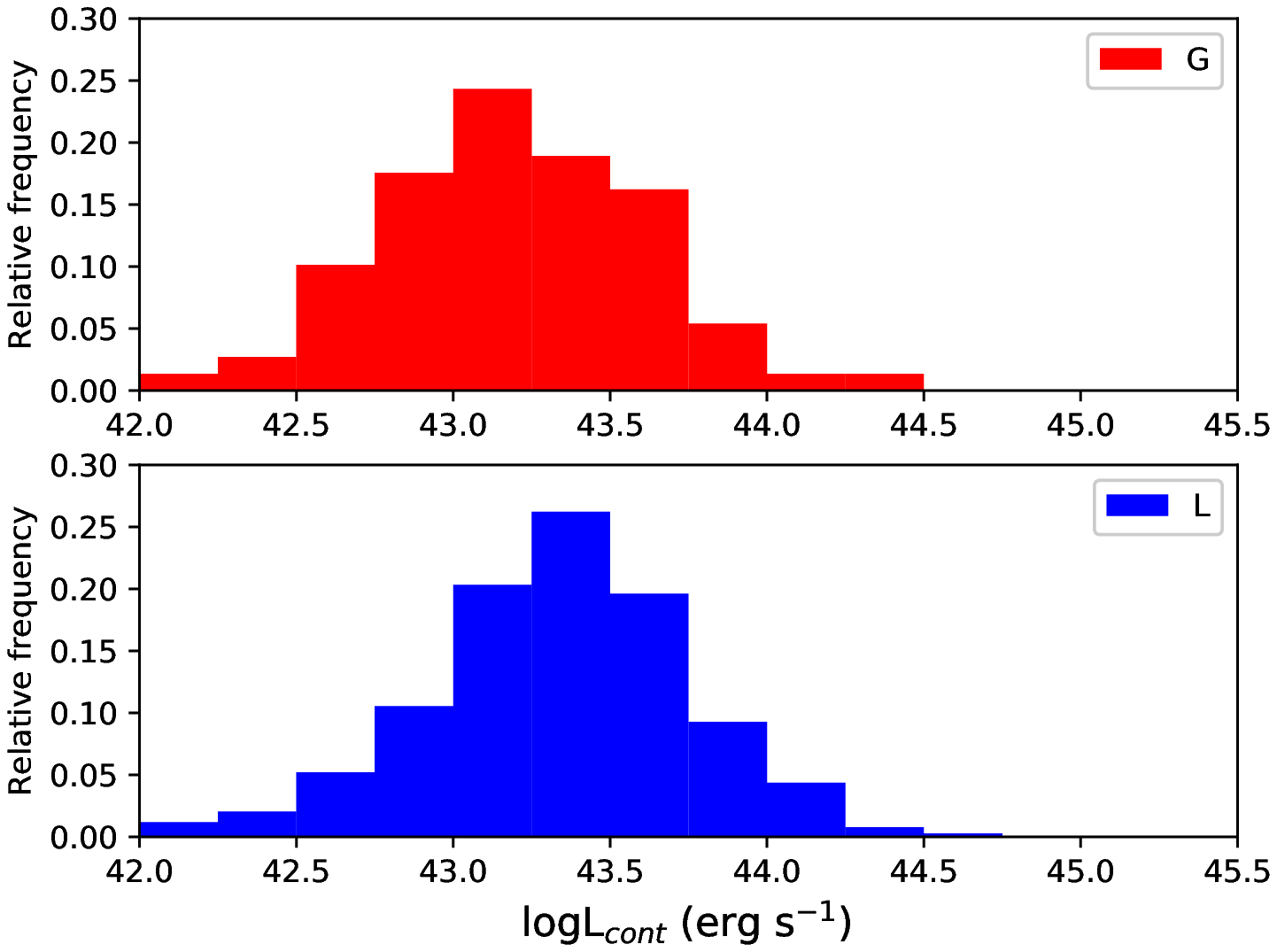}
\includegraphics[width=0.5\hsize,clip=]{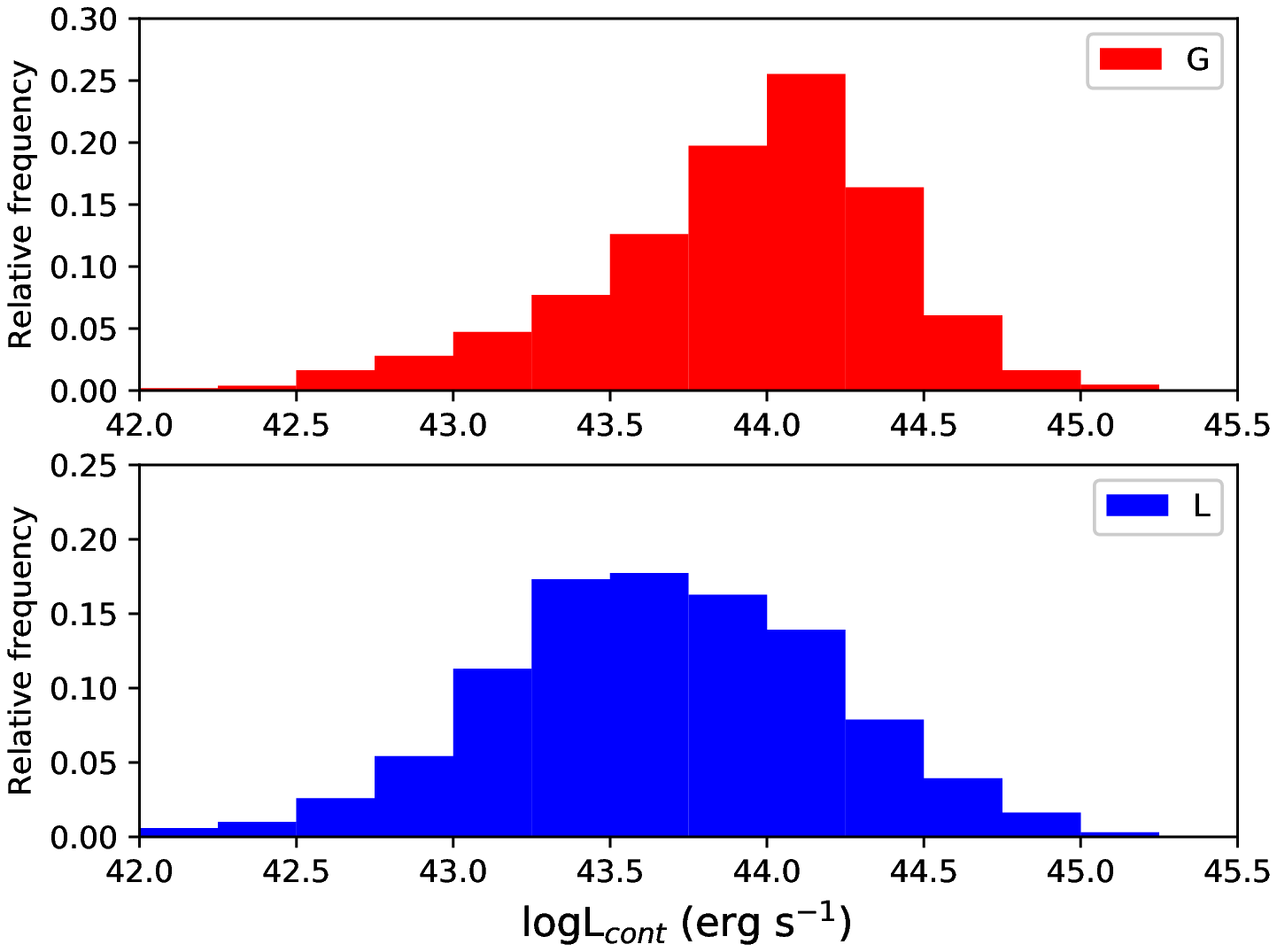}
\end{minipage}
\caption{Distribution of the monochromatic continuum luminosity at 5100 \AA\ of the two samples. Colors as in Fig.~\ref{fig:histo_z} and panels as in Fig.~\ref{fig:histo_bh}.}
\label{fig:histo_cont}
\end{figure}

\begin{figure}[!h]
\begin{minipage}[t]{\textwidth}
\includegraphics[width=0.5\hsize,clip=]{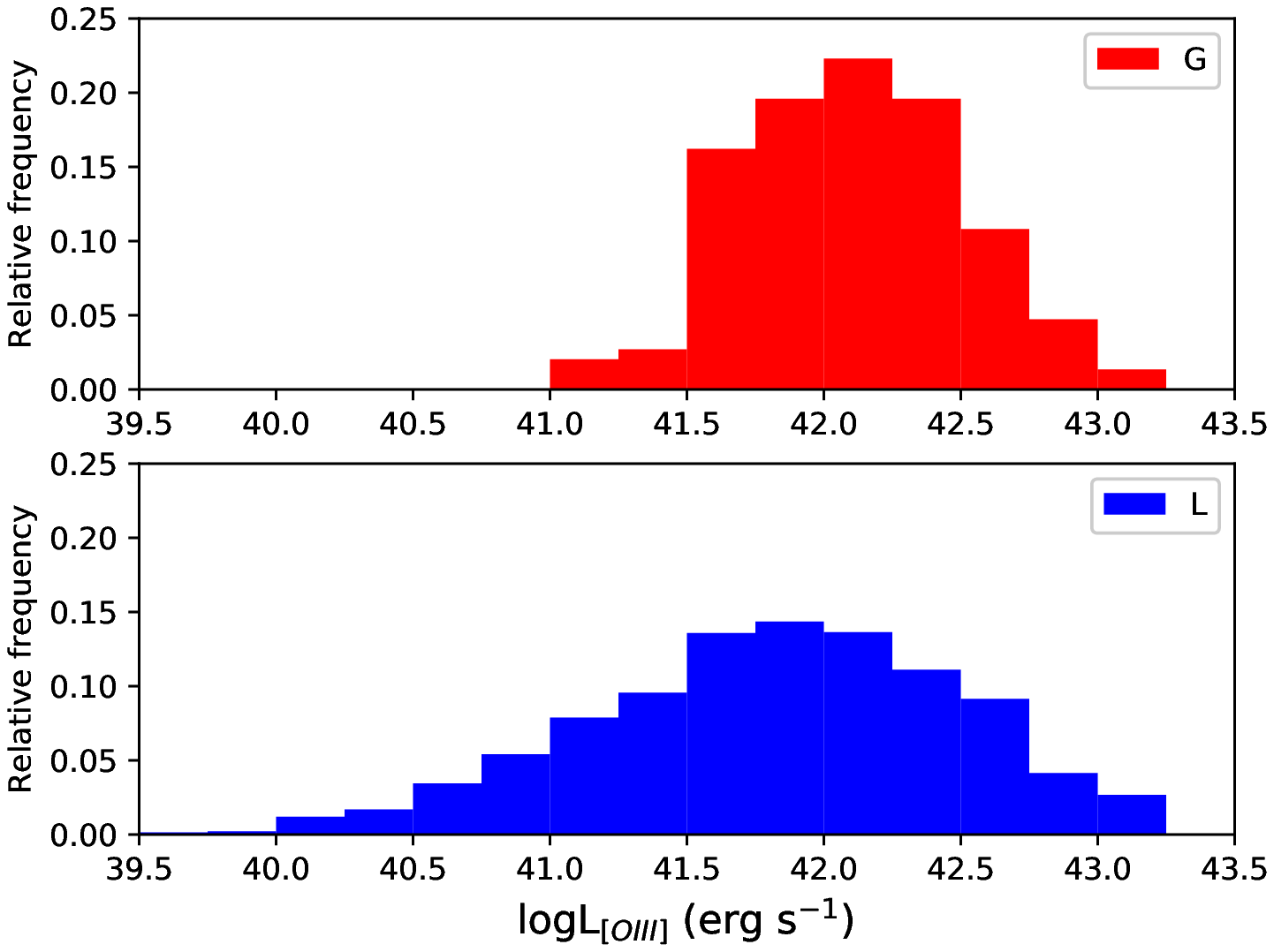}
\includegraphics[width=0.5\hsize,clip=]{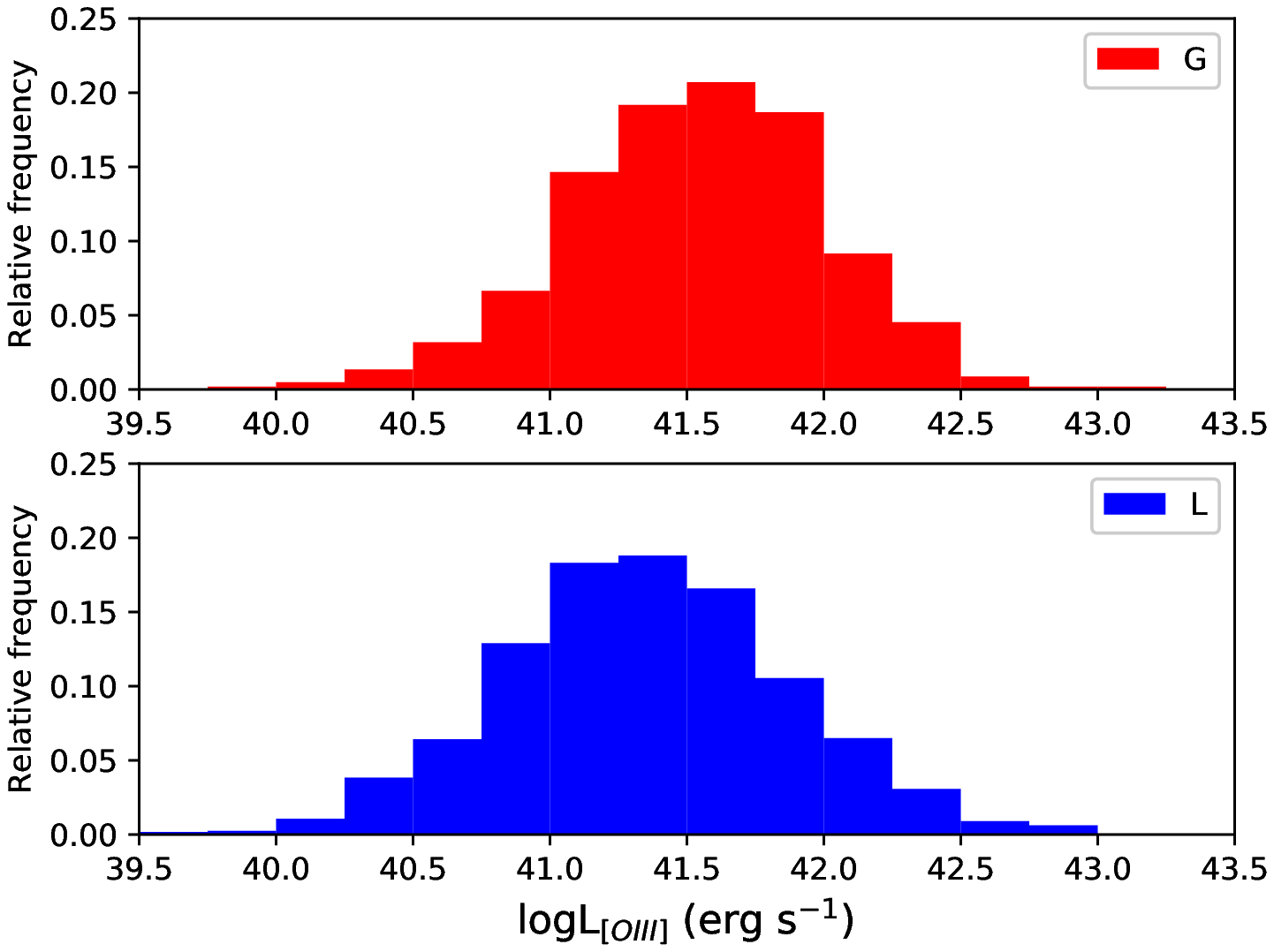}
\end{minipage}
\caption{Distribution of [O III] luminosity of the two samples. Colors as in Fig.~\ref{fig:histo_z} and panels as in Fig.~\ref{fig:histo_bh}.}
\label{fig:histo_o3}
\end{figure}

\section{Physical properties of the sources}
\label{sec:results}
As mentioned above, the physical quantities we extracted from R17 are black hole mass, Eddington ratio, relative Fe II strength (R4570), monochromatic continuum luminosity at 5100 \AA, and [O III] integrated luminosity. Their mean and median values, and the standard deviation of each sample are shown in Table~\ref{tab:stats}. The distributions of each physical property for both samples are shown in Figs.~\ref{fig:histo_bh},~\ref{fig:histo_edd},~\ref{fig:histo_fe},~\ref{fig:histo_cont}, and~\ref{fig:histo_o3}. All left panels refer only to sources with z$<$0.25, while the right panel shows the entire sample up to z$=$0.8. \par
The black hole mass in R17 was estimated under the assumption of virialized gas, with:
\begin{equation}
M_{BH} = f \frac{v^2 R_{BLR}}{G} \; .
\end{equation}
The FWHM(H$\beta$) broad component was used as a proxy for the rotational velocity of the gas surrounding the black hole. They assumed that the broad component can be always represented with a Lorentzian function. Therefore, they decomposed the H$\beta$ profile using a Lorentzian to represent the broad component, and a Gaussian to represent the narrow component. The BLR radius was calculated from the $\lambda$5100 \AA\ continuum luminosity with the relation by \citet{Bentz13}. The factor $f$, which is dependent on the BLR geometry, was fixed to 3/4 assuming a spherical distribution of clouds. The use of FWHM in black hole mass calculation is not ideal, since it has been shown that the second-order moment of the line is a better proxy for the velocity and less dependent on BLR geometry \citep{Peterson11, Foschini15, Berton15a, Peterson18}. However, since our aim is to compare the two distributions and not to obtain precise mass values, the use of FWHM is acceptable. The K-S test rejects the null hypothesis at high confidence level both at low redshift and in the whole sample. The difference however is relatively small up to z$=$0.25, while it is remarkably large when the whole sample is considered. In both cases, mean and median mass are lower among L sources, but the mass of G sources at high redshift is significantly larger than those at low $z$. \par
The Eddington ratio was calculated, following R17, as the ratio between the bolometric luminosity and the Eddington luminosity. For the bolometric luminosity they assumed it to be nine times the monochromatic luminosity at $\lambda$5100 \AA. The Eddington luminosity was calculated as L$_{Edd} = 1.3\times10^{38}$ M$_{\rm BH}$/M$_\odot$, where M$_{\rm BH}$ is the black hole mass. The distributions are rather different, but at z$<$0.25 this difference seems to be more significant. Given the shape of the distributions, seen in Fig.~\ref{fig:histo_edd}, we also carried out the A-D test to understand whether the tails of the two distributions are also statistically different. The test confirms the difference both at low redshift and for the whole sample, rejecting the null hypothesis at a 99\% confidence level. Indeed, it seems that the Eddington ratio is higher in L sources, and this is particularly evident at low redshift. At high redshift, the median values of the two samples are the same. However, there is a much larger number of high Eddington objects in the L sample than in the G sample, and this is in agreement with the larger mean value for this parameter in L sources, and with the result of the A-D test. \par
The Fe II multiplets flux was measured by R17 using the template provided by \citet{Kovacevic10}, and later divided by the H$\beta$ flux to obtain the R4570 parameter. The two distributions are not statistically different below z=0.25. This result is confirmed both by the K-S test and by the A-D test. Both, in fact, do not reject the null hypothesis. When instead the whole samples are compared, the null hypothesis is rejected by both tests at 99\% confidence level, suggesting that R4570 is in fact different between G and L sources. In particular, it seems that R4570 is slighly larger in L sources than in G sources. \par
The monochromatic luminosity of the continuum at 5100 \AA\ was measured directly on the spectrum. The K-S test allows us to reject the null hypothesis in both redshift intervals. However, the behavior radically changes with the redshift. While up to z=0.25 on average L sources are brighter than G sources, when high-z objects are considered the result is the opposite, with G sources becoming brighter. \par
Finally, the [O III] flux was measured by fitting the $\lambda\lambda$4959,5007 lines with two Gaussians each, one to represent the core component, and the second to reproduce the wing \citep[e.g., see][]{Sulentic00, Marziani03, Komossa08}. Both in the low redshift domain and in the whole sample, the K-S test allows us to reject the null hypothesis, suggesting that the two sample do not originate from the same population of sources. The [O III] luminosity is indeed larger in G sources than in L sources. The difference becomes more evident when sources at high redshift are included. This result is particularly interesting in the light of the behavior found in the continuum luminosity, where the ionizing photons originate. While at high-z an increase in continuum luminosity corresponds to an increased [O III] luminosity, as expected, at low-z this does not happen, suggesting that another factor contributes to the larger [O III] luminosity observed in G sources (see next Section). \par


\section{Discussion}
As shown in Sect.~\ref{sec:results}, basically all physical properties of Gaussian and Lorentzian NLS1s are different. In the following we will discuss them one by one, and in the end provide a speculative picture regarding the evolution of these objects. \par
First of all, out of 3933 sources, approximately 2894 ($\sim$74\%) can be better reproduced by a Lorentzian profile, while the remaining $\sim$26\% have a Gaussian line profile. The fraction of G sources is larger than what found by other studies. For example, \citet{Cracco16} found 96\% of the profiles to be Lorentzian. However, as discussed above, we found that the redshift distribution of Gaussian profiles is concentrated at higher redshifts than Lorentzian, therefore a selection effect may be present. Indeed, when the sample is limited to z=0.25, the percentage of L sources is $\sim$90\%, rather close to what was found in other works. \par
The different line profiles can be interpreted in terms of BLR geometry. As shown by \citet{Gaskell09}, the BLR geometry in AGN seems to have a bird's nest appearence with velocity components
\begin{equation}
v_{keplerian} > v_{turbulent} \ga v_{inflow} \; .
\end{equation}
The presence of Keplerian motion was remarkably showed with VLTI observations of 3C 273 \citep{Gravity18}, and several authors \citep[e.g.,][]{Kollatschny11, Goad12, Kollatschny13} claimed that the Gaussian profile originates when the BLR motion is dominated by a Keplerian rotation. However, in NLS1s the relation may be different, with 
\begin{equation}
v_{keplerian} \sim v_{turbulent} \; ,
\end{equation}
while nothing can be said (yet) about the inflow velocity. Lorentzian profiles seem in fact to be associated with microturbulence \citep{Goad12}, possibly pointing out that this kind of motion is significantly stronger in NLS1s. Such turbulent motion is likely associated with the vertical structure of the BLR, since it is due to random motion of the clouds above and below the equatorial plane \citep{Kollatschny13a}. Indeed, a prominent fountain-like vertical structure has been also observed in AGN with Lorentzian line profiles of the Mg II line \citep{Popovic19}. All of this may indicate that the BLR geometry in NLS1s is less flattened than in other AGN, in agreement with the independence of line width from AGN inclination \citep{Vietri18}. 

\subsection{Black hole mass}
Black hole mass has always been the most discussed property of NLS1s, in particular after the discovery of $\gamma$-ray emission from those with a beamed relativistic jet \citep{Abdo09a, Abdo09c}. The presence of relativistic jets in these low-mass AGN was indeed a shocking discovery, since it contradicted the well-established paradigm according to which only high mass black holes are able to launch a relativistic jet \citep{Laor00}. Some authors, indeed, tried to reconcile NLS1s, and extreme accretors in general, with the paradigm \citep{Decarli08, Chiaberge11}. However, reverberation mapping \citep{Peterson18} confirmed that NLS1s are characterized by a black hole with a mass lower than 10$^8$ M$_\odot$. Furthermore, the low jet power observed in jetted NLS1s is likely associated with their low black hole mass \citep{Foschini14, Foschini15}. Jet power, indeed, scales non-linearly with black hole mass \citep{Heinz03}. For the same reason, the radio luminosity of beamed jetted NLS1s \citep{Foschini17} is lower with respect to other blazars, suggesting that they may be the low-luminosity tail of the blazar population \citep{Berton16c}. Last but not least the fact that, unlike other AGN, NLS1s are typically hosted in disk galaxies \citep[both non-jetted and jetted, see][]{Crenshaw03, Deo06, Orbandexivry11, Mathur12, Kotilainen16, Olguiniglesias17, Jarvela18, Berton19a}, suggests that their black hole mass is lower than that of BLS1s, since the black hole mass tends to scale with the host galaxy morphology \citep{Kormendy13}. \par
Our findings indicate that black hole masses change within the NLS1 population. The L and G distributions, both at low redshift and among the whole sample, are different from each other. This difference seems to be more prominent at large $z$. A selection effect could be present since the most luminous sources, seen at larger distances, may have a larger black hole mass and, as mentioned above, they preferentially belong to G sample. However, the presence of this difference at low $z$ seems to indicate that it is not entirely a selection effect. Furthermore, this result is somewhat expected. The black hole mass is linearly dependent on the FWHM of \Hb\ and a Lorentzian profile, for simple geometrical reasons, has a lower FWHM than a Gaussian profile with the same flux. If the \Hb\ flux does not change, therefore, the derived black hole mass will be lower in Lorentzian sources (which in turn gives a higher Eddington ratio, a feature we also observe). \par
We must stress that such different behavior is not a bias: the line profile reflects a difference in the physical properties of the sources, therefore the different black hole mass distribution is a real effect. Furthermore, the black hole mass calculation we adopted in this work is the same for both line profiles. As mentioned above, the black hole mass we used are the same estimated by R17, who decomposed the H$\beta$ profile always using a Lorentzian function to represent the broad component and a Gaussian to reproduce the narrow component, regardless of the function which better represents the whole line profile. Assuming that the narrow component is the same in both cases, a Lorentzian broad component would yield to a lower FWHM with respect to a Gaussian broad component, thus slightly underestimating the black hole mass in the Gaussian sample. As a result, the intrinsic separation between the mass distributions of L and G sources should be even larger. \par

\subsection{Eddington ratio, continuum and [O III] luminosity}
It is widely known that NLS1s are extreme accretors, typically showing a rather high Eddington ratio \citep{Boroson92, Marziani18b, Marziani19}. This phenomenon is believed to be the driver of the EV1 properties of AGN, and it actually accounts for many attributes of NLS1s. Strong disk winds are often observed in these sources \citep[e.g.,][]{Jin17a, Jin17b, Gallo19}, and the high Eddington ratio could also be responsible for the bulk outflowing motion often observed in their NLR \citep{Zamanov02, Marziani03, Komossa08}. \par
In our samples, we found that Lorentzian sources have a systematically different Eddington ratio, usually higher with respect to Gaussian sources. The mean and median values of Eddington ratio of the whole sample are higher than the mean and median of low redshift objects. Since black hole masses tend also to be larger when the entire sample is considered, this indicates that the bolometric luminosity of sources at high redshift is larger. This is expected, since low-luminosity sources at high $z$ are simply not visible. Therefore, the effect of redshift is very prominent on the results we obtained for the whole sample. \par
What we observe in the redshift-limited sample is way more interesting. The difference between the two distribution is very large, and significantly larger than the difference observed between the two black hole mass distributions. This indicates that the bolometric luminosity of G sources and L sources are also different at low z, as shown by the different distributions of continuum luminosity\footnote{We remind that the bolometric luminosity was estimated by multiplying the monochromatic luminosity times a constant value.}. In particular, it seems that at low redshift L sources are brighter than G sources. Indeed, as found from the continuum luminosity, in our low-z sample the median value of bolometric luminosity of G sources is 1.4$\times$10$^{44}$ \ergs, while in L sources the median is 2.0$\times$10$^{44}$ \ergs. Even if the difference is not so pronounced, this is an interesting result since it is in contradiction with what we observe in the [O III] luminosity. The G sources, indeed, are brighter than L sources (1.3$\times$10$^{42}$ \ergs\ for G sources and 6.9$\times$10$^{41}$ \ergs\ for L sources), but the brighter continuum in L sources should indeed produce brighter high ionization lines, such as [O III], within the same class of objects. The difference in [O III] distribution also rules out the possibility that the difference between L and G sources is inclination. Since [O III] luminosity is an isotropic property, if L and G sources were seen at different angles their [O III] luminosities should be the same.

\subsection{Iron and R4570}
The origin of iron in NLS1s is still unclear. However, it is well known that iron originates in supernova explosions, particularly from Ia and core-collapse supernovae (Ib, Ic, II). Authors already suggested that the origin of metals in AGN can be connected with starburst events \citep{Heller94, Collin99}. NLS1s, in particular, are well known to be objects with an enhanced star formation rate with respect to BLS1s \citep{Sani10, Caccianiga15}. A high star formation rate can naturally lead to the explosion of several core-collapse supernovae in addition to Ia SNe, thus increasing the metallicity of these AGN \citep{Chen09, Million11}. When galaxy-galaxy interaction is ongoing, as observed in the hosts of several jetted NLS1s \citep{Anton08, Dammando18, Jarvela18, Berton19a}, star formation events and supernovae explosions may be even more common. For example, in the case of IRAS 20181-2244, the star formation rate was estimated from the near infrared to be close to 300 M$_\odot$ yr$^{-1}$ \citep{Caccianiga15}, and its host galaxy is clearly undergoing a major merging event \citep{Berton19a}. These events may trigger a large number of core-collapse supernovae explosions in a rather short timescale. Additionally, given the typical velocity of supernovae ejecta (5000 \kms), the products of the explosion can cross a Milky Way-sized galaxy in a few $10^6$ years, a timescale possibly shorter than a typical NLS1 activity event ($10^8$ years, \citealp{Komossa18}). If the star formation is circumnuclear, as pointed out by \citet{Sani10}, the timescale to feed iron and other metals into the nuclear region is even shorter. \par
The distributions of R4570 are statistically different when the whole sample is considered, but at low $z$ the R4570 seems identical in L and G sources. It is worth noting that R4570 does not depend only on Fe II strength, but also on the H$\beta$ flux. Therefore, a lower R4570 does not necessarily reflect on the Fe II flux directly \citep{Cracco16}. In the case of our low-z sample, in analogy with what happens for [O III], the H$\beta$ luminosity is higher in the G sample than in the L sample (median logarithmic values 42.45 vs 42.18 for G and L, respectively). If the R4570 in L and G sources is approximately the same, this result implies that the Fe II strength is higher in G sources. The same result is found when using the whole samples. In this case the ratio R4570 is generally lower in G sources. As before, H$\beta$ is instead significantly brighter in G sources, therefore the Fe II strength is even higher when high-z sources are included in the sample. 

\subsection{A complete picture? EV1 and intraclass evolution}
\begin{figure}[!t]
\begin{minipage}[t]{\textwidth}
\includegraphics[width=0.5\hsize,clip=]{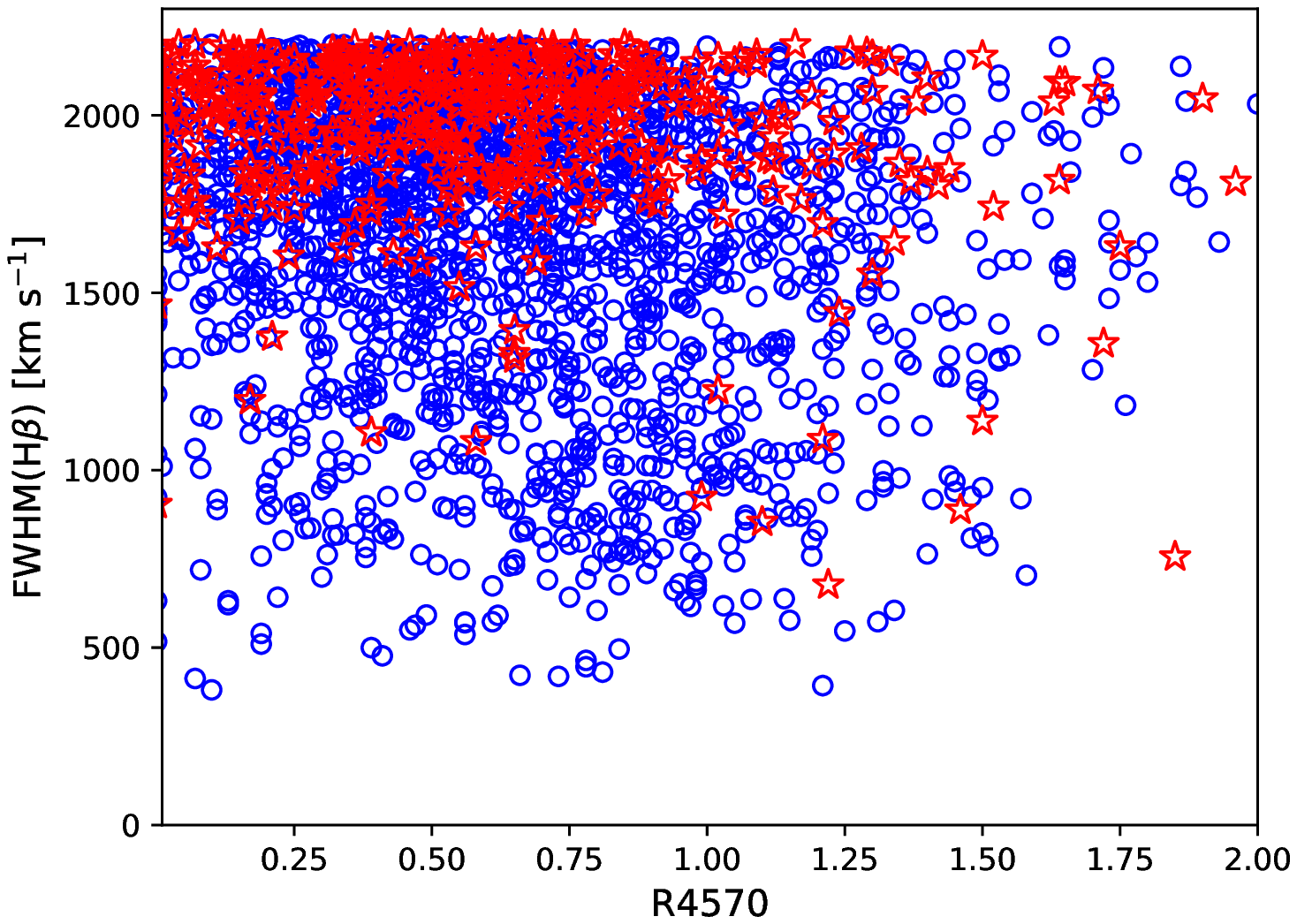}
\includegraphics[width=0.5\hsize,clip=]{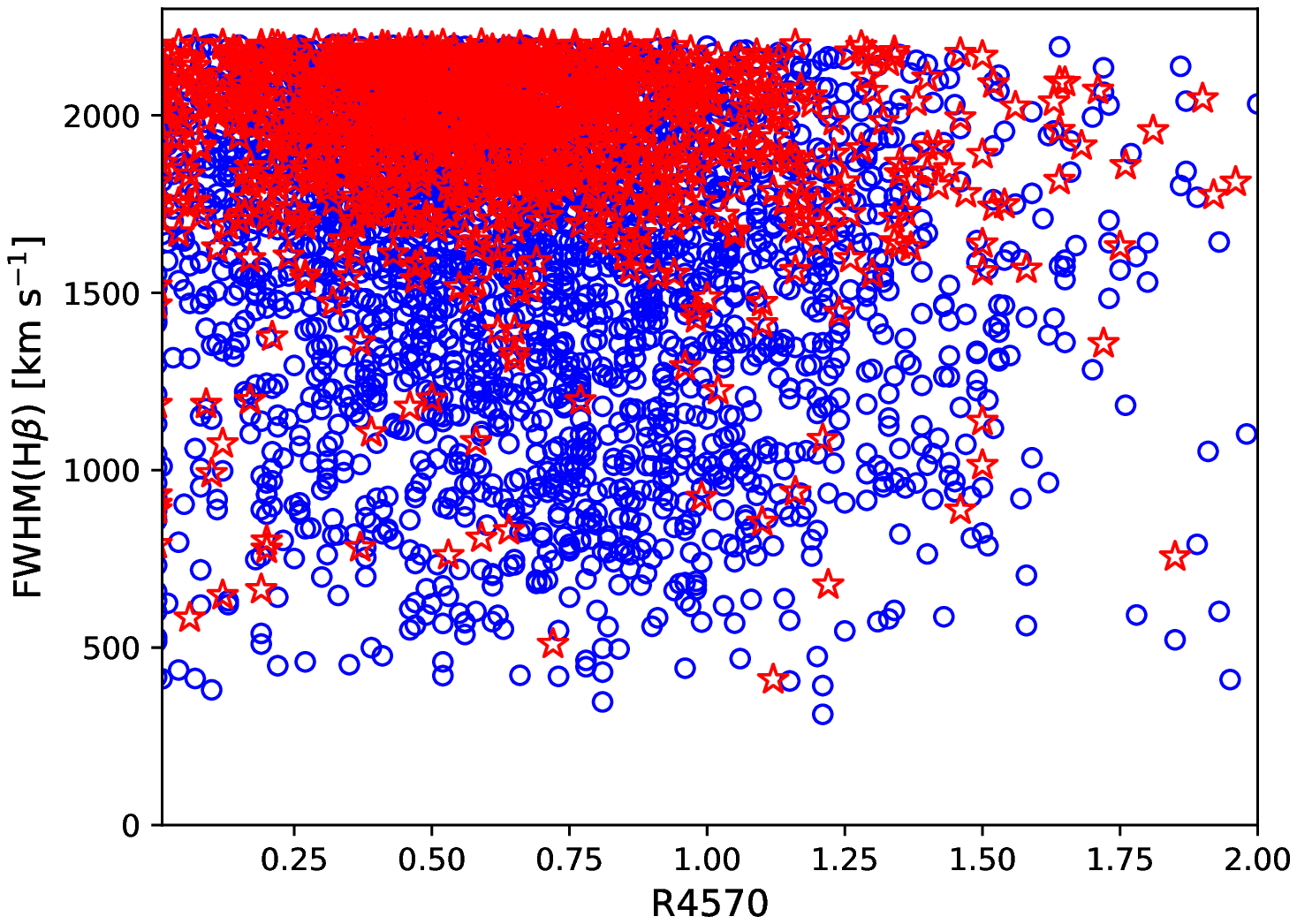}
\end{minipage}
\caption{Position of our sources on the EV1 plane. The x-axis is the R4570, the y-axis is the FWHM(H$\beta$). Colors as in Fig.~\ref{fig:histo_z} and panels as in Fig.~\ref{fig:histo_bh}.}
\label{fig:ev1}
\end{figure}
To have a better understanding of the previous results, we plotted our sources on the EV1 plane. Although in general the y-axis of EV1 is the FWHM of H$\beta$ broad component, in our case the H$\beta$ line is well represented by a single component, therefore we used the total FWHM of the line. What is shown in Fig.~\ref{fig:ev1} is extremely interesting. While L sources are basically spread on the whole diagram, the bulk of G sources is located close to the FWHM threshold of NLS1s, and on the left part of the diagram. This is the region where the vertical branch of the EV1 begins. This may suggest that G sources act like a ``bridge", connecting Lorentzian sources of the so-called population A (FWHM$<$4000 \kms, \citealp{Sulentic00}) with the mostly Gaussian sources that belong to population B (FWHM$>$4000 \kms). \par
As already mentioned, many authors suggested that the driver of EV1 is the Eddington ratio \citep{Boroson92}. This idea is consistent with what we see in our plot, since G sources are located on the upper part of the diagram, where the Eddington ratio is supposed to decrease. However, some authors suggested that the EV1 represents the equivalent of a main sequence for quasars \citep{Sulentic15}, and that the position of the sources on this sequence may be connected to their age \citep{Fraixburnet17a, Fraixburnet17b}. Indeed, the black hole mass could be used as a sort of an ``arrow of time". Since black hole mass can only grow, young sources must have lower masses than old sources. On the EV1, this would mean that population A sources are the progenitors of population B objects, with NLS1s being the youngest of all. Furthermore, given the different distributions of G and L sources on the EV1, and the fact that population B sources are largely dominated by Gaussian line profiles, we hypothesize that intraclass evolution could occur within the NLS1 population, and that L sources are the progenitors of G sources. It is already well known, in fact, that there is no sharp transition between NLS1s and BLS1s, but their properties form a continuous distribution \citep[e.g., ][]{Cracco16}. It is therefore natural to expect the existence of an intermediate class of sources, which constitutes a link between the youngest NLS1s and their older evolved counterparts. \par
This evolutionary scenario could account for our previous results. The black hole mass increases going from L to G sources, in agreement with what mentioned before. If L sources are recently switched-on AGN, it is possible that they are surrounded by a dense environment consisting of gas clouds, dust, and stars. In these initial phases the accreting material is more abundant, which could account for the higher Eddington ratio. When the activity goes on for a long time, the Eddington ratio slightly decreases, while the AGN approaches the BLS1 phase and moves on the EV1 toward population B. We highlight that this model does not require a continuous single accretion episode. Separate activity episodes, possibly due to disk instabilities \citep{Czerny09}, do not affect this hypothesis. The [O III] luminosity could be due to a change in the BLR geometry, reflected by the different line profiles. As mentioned above, L sources may have a non-negligible vertical structure, which could increase the covering factor for the NLR. The ionizing continuum produced by the accretion disk that could reach the NLR may be absorbed by a more spherical BLR, accounting for weaker high ionization lines. With time, disk winds and radiation pressure may blow away the turbulent clouds, leaving a BLR dominated by Keplerian motion, and decreasing the covering factor for the NLR. \par
Finally, Fe II strength may also be included into this scenario, if we assume that the Fe II abundance is connected to the star formation rate. AGN activity is known to heat the gas and suppress star formation \citep{Morganti17}. Therefore, supernova events may be less common in sources where the AGN activity has been ongoing for a long time. However, if NLS1s evolve from Lorentzian objects to Gaussian objects, we should observe an increasing amount of iron with increasing age. All the iron atoms, indeed, are formed during the early phase of the AGN life, and this behavior is essentially what we observe among our samples. \par
An interesting example that may fit into our picture is that of Mrk 783. This NLS1 was not included in our study, since its H$\beta$ profile cannot be reproduced by a single component \citep{Berton15a, Congiu17, Congiu17b}. However, when the narrow component is subtracted, its line profile can be classified as Gaussian. In radio this NLS1 shows a prominent diffuse radio emission on kpc-scale, which can be interpreted in terms of relativistic jet precession or even as a relic emission. Therefore, despite being an NLS1, it is unlikely to be a very young source since, unlike most jetted NLS1s \citep{Berton18a}, it already had the time to develop a prominent diffuse emission. Its other properties are reminiscent of those of Gaussian sources. The black hole mass is in the higher end of NLS1s distribution (4.3$\times$10$^7$ M$_\odot$, \citealp{Berton15a}), its Eddington ratio is quite low (0.11, \citealp{Berton15a}), and its [O III] line is quite strong with respect to H$\beta$ (R5007 $\sim$ 2.9). Only the iron is slightly different from the typical properties of the Gaussian sample, since the iron is basically not present with the only exception of a few high ionization lines (Congiu et al., in prep.). However, its  Mrk 783, in conclusion, could be a good example for this scenario, since its properties seem to fit this evolutionary picture. \par

\section{Summary}
In this work we analyzed the physical properties of two subsamples of NLS1 galaxies extracted from the SDSS. We divided them into Lorentzian and Gaussian sources according to the profile that better represents their H$\beta$ line. We later introduced a redshift limit at z=0.25 in the sample, in order to study both low-redshift objects and the sample as a whole. Our main result is that most of their physical properties seem to differ. From Lorentzian to Gaussian sources the black hole mass tends to increase, the Eddington ratio decreases, the Fe II strength decreases, and the [O III] luminosity increases. \par
We try to fit these differences into the context of the quasar main sequence and the EV1, suggesting that Lorentzian sources are the progenitors of Gaussian sources. This groundbreaking scenario seems to account  for most of the observed results. In particular, the increased black hole mass may be used as an arrow of time which is representative of AGN evolution. If this hypothesis is correct, AGN may be born as NLS1s with Lorentzian line profiles, later evolve to NLS1s with a Gaussian profile, and eventually move in the EV1 region where population B sources are. \par
This model is still rather speculative, and needs further confirmation. A more careful analysis of the line profiles on large samples of NLS1s is needed in order to better separate Gaussian and Lorentzian sources. Furthermore, this study should be extended to the entire population A of the EV1, to see if this trend can be confirmed, and eventually to all type 1 AGN. An upcoming paper will be dedicated to the careful analysis of NLS1 spectra, and will provide a more complete view on this important issue.


\acknowledgements
MB would like to warmly thank the organizers of the 12th Serbian Conference on Spectral Line Shapes in Astrophysics for the interesting and fruitful meeting. The authors are grateful to P. Marziani, W. Kollatschny, E. Bon, S. Ciroi, and L. Foschini for helpful suggestions and discussions. This research has made use of the NASA/IPAC Extragalactic Database (NED) which is operated by the Jet Propulsion Laboratory, California Institute of Technology, under contract with the National Aeronautics and  Space Admistration. Funding for the Sloan Digital Sky Survey has been provided by the Alfred P. Sloan Foundation, and the U.S. Department of Energy Office of Science. The SDSS web site is \texttt{http://www.sdss.org}. SDSS-III is managed by the Astrophysical Research Consortium for the Participating Institutions of the SDSS-III Collaboration including the University of Arizona, the Brazilian Participation Group, Brookhaven National Laboratory, Carnegie Mellon University, University of Florida, the French Participation Group, the German Participation Group, Harvard University, the Instituto de Astrofisica de Canarias, the Michigan State/Notre Dame/JINA Participation Group, Johns Hopkins University, Lawrence Berkeley National Laboratory, Max Planck Institute for Astrophysics, Max Planck Institute for Extraterrestrial Physics, New Mexico State University, University of Portsmouth, Princeton University, the Spanish Participation Group, University of Tokyo, University of Utah, Vanderbilt University, University of Virginia, University of Washington, and Yale University.

\bibliography{/home/berton/Desktop/Paper/biblio}

%

\clearpage
\end{document}